\documentclass{article}

\usepackage[preprint]{neurips_2026}

\usepackage[utf8]{inputenc}
\usepackage[T1]{fontenc}
\usepackage{hyperref}
\usepackage{url}
\usepackage{booktabs}
\usepackage{amsfonts}
\usepackage{amsmath}
\usepackage{nicefrac}
\usepackage{microtype}
\usepackage{graphicx}
\usepackage{xcolor}
\usepackage{subcaption}
\usepackage{multirow}

\graphicspath{{figures/}}

\title{Non-Monotonic Safety Alignment Across Gemma Generations: \\
A Quality-Diversity Red-Teaming Study}

\author{%
  Subhadip Mitra\thanks{This paper builds on the quality-diversity adversarial framework introduced in our prior work~\citep{mitra2026qd}, which appeared at the ICLR 2026 Workshop on Agents in the Wild (AIWILD).}\\
  Research Lead, Rota Labs\\
  \texttt{subhadip@rotalabs.ai}\\
}

\begin{document}

\maketitle

\noindent\fbox{\parbox{\dimexpr\linewidth-2\fboxsep-2\fboxrule\relax}{\small
\textbf{Note on v2 (November 2026).} This version counts an attack as successful only when the
judge labels the response both non-refusing \emph{and} harmful; v1 counted every non-refusal.
Rates are therefore lower, and v1's figures are reported as the \emph{non-refusal rate}. With
three seeds per model, this version makes no significance claims. The qualitative ordering is
unchanged: Gemma~3 is the most vulnerable generation and Gemma~4 the least. Details:
\S\ref{sec:method:qd}, \S\ref{sec:method:scaling}.}}
\vspace{0.75em}

\begin{abstract}
Safety alignment in LLMs need not improve monotonically across model
generations. We study four generations of Google's Gemma family
(7B--31B), using quality-diversity evolution (MAP-Elites) as an automated
red-teaming probe. An attack counts as successful when an 8B LLM judge
labels the response as non-refusing and harmful. Gemma~3 (12B) has the
highest attack success rate (ASR): $33.7\% \pm 1.3\%$ (mean $\pm$ std,
3 seeds), against $19.0\% \pm 10.6\%$ for its predecessor Gemma~2 and
$12.3\% \pm 4.2\%$ for its successor Gemma~4. Every Gemma~3 seed exceeds
every Gemma~2 and Gemma~4 seed. With three seeds this is the strongest
separation a permutation test can show ($p = 0.10$), so we report the
pattern as directional. Harmful successes rarely transfer between
generations: replaying evolved archives on other generations yields
1.4--7.9\% ASR, lowest on Gemma~4 (1.4--2.8\%). Category profiles shift
sharply. Cybercrime is the most exploited category in Gemma~1--3
(47--83\%) but falls to 2\% in Gemma~4, while misinformation rises from
17\% (Gemma~2) to 80\% (Gemma~3) and is still 53\% in Gemma~4. A
second-judge audit (Llama-3.1-70B) confirms 74\% of the 8B judge's
harmful labels. We also report the non-refusal rate, the metric used in
v1 of this paper. It shows the same ordering at 2--3$\times$ higher
levels (34--69\%), and overstates harmful transfer by 6--16$\times$.
All experiments use 3 random seeds with a unified self-hosted judge;
code and artifacts are available at
\url{https://github.com/bassrehab/red-queen}.
\end{abstract}

\section{Introduction}
\label{sec:intro}

Large language model (LLM) safety alignment has improved rapidly, but
measuring that improvement rigorously remains an open challenge.
Static benchmarks such as HarmBench~\cite{mazeika2024harmbench} and
AdvBench~\cite{zou2023universal} provide point-in-time snapshots but
do not capture how an adaptive adversary interacts with evolving defenses.

A fundamental question is whether safety improvements \emph{generalize}:
when a model developer patches vulnerabilities between releases, do old
attacks stop working (indicating targeted fixes), or does the model become
broadly more robust? Answering this requires a methodology for
\emph{differential safety probing}: systematically comparing safety
surfaces across model versions with a consistent adversarial process.

In this paper, we address two specific questions:
\begin{enumerate}
    \item \textbf{Cross-generational transfer:} Do adversarial attacks
          evolved against an earlier model generation transfer to later
          generations of the same family?
    \item \textbf{Non-monotonic robustness:} Does safety alignment
          improve consistently across model generations, or can it
          regress?
\end{enumerate}

We study these questions using four generations of Google's Gemma model
family (7B, 9B, 12B, 31B parameters), spanning releases from 2024 to 2026.
Our probe is a quality-diversity (QD) evolutionary algorithm (MAP-Elites)
that discovers diverse jailbreak strategies across seven harm categories
from the HarmBench taxonomy. By evolving attack archives against each model
with 3 independent seeds and replaying them across generations, we
construct a transfer matrix with uncertainty estimates that reveals which
vulnerabilities persist and which are patched.

\paragraph{Contributions.}
\begin{itemize}
    \item A QD-evolution methodology for \emph{differential safety probing}
          that enables systematic, reproducible comparison of safety surfaces
          across model versions (\S\ref{sec:method}).
    \item A cross-generational transfer analysis across four Gemma
          generations. It shows that attacks eliciting harmful content
          rarely transfer between generations (1.4--7.9\%), and that
          Gemma~4 receives the least transfer
          (\S\ref{sec:results:transfer}).
    \item Evidence, directional with three seeds, that adversarial
          robustness does not improve monotonically across model
          generations within a single family. Gemma~3 has the highest ASR
          in every seed, coinciding with an architectural transition
          (\S\ref{sec:results:scaling}).
    \item Category-level analysis showing that most harmful successes
          come from cybercrime and misinformation, with category profiles
          that change sharply between generations
          (\S\ref{sec:results:baseline}).
    \item A measurement note: in our data, counting non-refusals as
          successes (the metric v1 reported as ASR) overstates harmful attack
          success by 2--3$\times$ for self-evaluation and 6--16$\times$
          for transfer (\S\ref{sec:discussion}).
\end{itemize}

\section{Related Work}
\label{sec:related}

\paragraph{Automated red-teaming.}
Recent work has developed automated methods for discovering LLM
vulnerabilities, including gradient-based attacks
(GCG;~\cite{zou2023universal}), LLM-based attackers
(PAIR~\cite{chao2023jailbreaking}; TAP~\cite{mehrotra2024tree}),
and template-based approaches (AutoDAN~\cite{liu2024autodan}).
These methods target individual models; none study how discovered
vulnerabilities transfer across model generations.

\paragraph{Safety benchmarks.}
HarmBench~\cite{mazeika2024harmbench} provides a standardized taxonomy
of harmful behaviors across seven categories.
StrongREJECT~\cite{souly2024strongreject} proposes improved evaluation
metrics. We use HarmBench as our behavior taxonomy and seed pool,
enabling category-level transfer analysis.

\paragraph{Quality-diversity optimization.}
MAP-Elites~\cite{mouret2015illuminating} maintains a grid of diverse,
high-performing solutions across behavioral dimensions. Prior work has
applied QD methods to game playing~\cite{fontaine2020covariance} and
robot design~\cite{cully2015robots}.
In prior work~\citep{mitra2026qd}, we established that MAP-Elites
discovers more diverse LLM vulnerabilities than standard evolutionary
search, in a single-model setting. We adopt that probe and extend it in
three directions: a cross-generational transfer replay protocol,
a four-generation longitudinal study with three seeds per model,
and evidence that safety alignment regresses non-monotonically
across Gemma generations.

\paragraph{Scaling laws.}
Scaling laws for LLM capabilities are well-established~\cite{kaplan2020scaling,
hoffmann2022training}. Whether safety properties follow similar predictable
scaling is largely unexplored. We provide evidence that adversarial
robustness does \emph{not} scale monotonically within a single model
family, a finding that contrasts with the regularity observed in
capability scaling.

\section{Methodology}
\label{sec:method}

\subsection{Quality-Diversity Adversarial Probing}
\label{sec:method:qd}

We use MAP-Elites to evolve a diverse archive of jailbreak attacks against
a target LLM. Each attack is characterized by a genome encoding:
\begin{itemize}
    \item \textbf{Attack strategy} (6 types): roleplay, encoding,
          authority impersonation, hypothetical framing, multi-turn, and
          direct jailbreak.
    \item \textbf{Encoding} (6 types): none, Base64, ROT13, leetspeak,
          pig latin, Unicode substitution.
    \item \textbf{Seed behavior}: drawn from HarmBench's 400 behaviors
          across 7 semantic categories.
\end{itemize}

The MAP-Elites archive is a 3D grid indexed by
(strategy $\times$ encoding $\times$ harm category), with each cell
storing the highest-fitness attack for that behavioral niche. Fitness is
evaluated by a self-hosted LLM judge (Llama-3.1-8B-Instruct) that
classifies each target response along three binary axes: refusal,
harmful content, and relevance. The resulting score ranges from 0
(refused) to 1 (harmful and relevant), with intermediate values for
non-refusing responses that are harmful but off-target (0.7), relevant
but not harmful (0.3), or neither (0.1). Evolution maximizes this
score. Using a single, self-hosted judge for \emph{all} phases
(evolution and transfer replay) ensures scoring consistency and full
reproducibility.

\paragraph{Success criterion.} An attack is \emph{successful} if the judge
labels the response as not refused and containing harmful content
(score $\geq 0.7$). ASR is the fraction of archive entries that are
successful. We separately report the \emph{non-refusal rate} (NRR): the
fraction of entries whose response is not labeled a refusal (score
$> 0$), regardless of harmfulness. Version~1 of this paper reported NRR
under the name ASR. The two differ substantially (\S\ref{sec:results}),
because many non-refusing responses are relevant but not harmful.

Each model is probed with 3 independent random seeds (42, 1337, 2718)
to estimate cross-seed variance. All evolution uses deterministic RNG
(ChaCha8, seeded) with ordered evaluation to ensure reproducibility
given the same seed.

\subsection{Cross-Generational Transfer Protocol}
\label{sec:method:transfer}

After evolving an archive $\mathcal{A}_{i,s}$ against model $M_i$ with
seed $s$, we \emph{replay} each archive against all other models
$M_j$ ($j \neq i$):
\begin{enumerate}
    \item For each entry in $\mathcal{A}_{i,s}$, send the attack prompt to
          $M_j$ and obtain the response.
    \item Score the response using the same LLM judge.
    \item Record per-entry and per-category transfer success, using the
          success criterion of \S\ref{sec:method:qd}.
\end{enumerate}

This produces a transfer tensor $T[i,j,s]$ where each entry is the
fraction of attacks from $\mathcal{A}_{i,s}$ that succeed against $M_j$.
We report the mean and standard deviation across seeds.
The diagonal $\bar{T}[i,i]$ is the mean self-evaluation success rate.
Off-diagonal entries reveal:
\begin{itemize}
    \item \textbf{Forward transfer} ($i < j$): old attacks on newer models.
          High values suggest safety improvements are superficial.
    \item \textbf{Backward transfer} ($i > j$): newer attacks on older
          models. High values suggest attacks exploit longstanding weaknesses.
\end{itemize}

\subsection{Robustness Across Generations}
\label{sec:method:scaling}

We plot robustness $R = 1 - \text{ASR}$ as a function of model
generation (with parameter count on the axis) to visualize the
trajectory of safety alignment within the Gemma family. We treat the
four generations as a case study.

\paragraph{Statistical scope.} With three seeds per model, a two-sided
permutation test comparing two models cannot yield $p < 0.10$. The
observed split and its mirror image are 2 of the $\binom{6}{3} = 20$
equally likely splits. We therefore report effect sizes, and whether
the per-seed values of two models are fully separated. We make no
claims of statistical significance. (Version~1 reported $p = 0.030$ and
$p = 0.032$ from a permutation routine whose strict floating-point
comparison dropped re-orderings of the observed split; those values are
artefacts.)

\section{Experimental Setup}
\label{sec:setup}

\subsection{Models}

\begin{table}[t]
\centering
\caption{Models evaluated. All served via vLLM~0.19.1 (bf16, temperature 0)
on a single NVIDIA H200 (141\,GB). Judge: self-hosted Llama-3.1-8B-Instruct
on NVIDIA RTX~4090 (24\,GB).}
\label{tab:models}
\begin{tabular}{llccc}
\toprule
\textbf{Model} & \textbf{Family} & \textbf{Params} & \textbf{Release} & \textbf{Architecture} \\
\midrule
Gemma 1 (7B-it)\textsuperscript{a}   & Gemma & 7B  & 2024 & Text-only \\
Gemma 2 (9B-it)\textsuperscript{b}   & Gemma & 9B  & 2024 & Text-only \\
Gemma 3 (12B-it)\textsuperscript{c}  & Gemma & 12B & 2025 & Multimodal \\
Gemma 4 (31B-it)\textsuperscript{d}  & Gemma & 31B & 2026 & Multimodal \\
\bottomrule
\end{tabular}
\\[2pt]
{\footnotesize
\textsuperscript{a}\texttt{google/gemma-7b-it}\quad
\textsuperscript{b}\texttt{google/gemma-2-9b-it}\quad
\textsuperscript{c}\texttt{google/gemma-3-12b-it}\quad
\textsuperscript{d}\texttt{google/gemma-4-31B-it}
}
\end{table}

We note that Gemma~3 and Gemma~4 introduced multimodal (vision-language)
capabilities, representing an architectural shift from the text-only
Gemma~1 and Gemma~2. This distinction is relevant to interpreting the
safety regression observed in Gemma~3 (\S\ref{sec:discussion}).

\subsection{Evolution Configuration}

MAP-Elites with $6 \times 6 \times 8$ grid (strategy $\times$ encoding
$\times$ category), yielding 288 behavioral niches. Batch size 50,
200 generations per model per seed, concurrency 5 parallel evaluations.
Mutation rate 0.3, crossover rate 0.7.
Seed pool: 400 HarmBench behaviors across 7 categories.
Three seeds per model (42, 1337, 2718) yield 12 evolution runs total.
Transfer replay: 36 source--target pairs (4 models $\times$ 3 other
targets $\times$ 3 seeds).

\subsection{Compute}

All Gemma models served sequentially on a single NVIDIA H200
(141\,GB HBM3e) via vLLM~0.19.1. LLM judge (Llama-3.1-8B-Instruct)
served on a dedicated NVIDIA RTX~4090 (24\,GB).
Total GPU time: approximately 60 hours across both machines
(12 evolution runs at $\sim$4 hours each, 36 transfer replays at
$\sim$8 minutes each, plus 4 raw-seed baselines and a 70B judge audit).

\section{Results}
\label{sec:results}

\subsection{Baseline Attack Success Rates}
\label{sec:results:baseline}

Table~\ref{tab:asr} presents attack success rates (ASR) across all
HarmBench categories for each Gemma generation, reported as mean $\pm$
standard deviation across 3 seeds. The most striking observation is
the \emph{non-monotonic} pattern. Overall ASR is $21.3\% \pm 3.7\%$
for Gemma~1 and $19.0\% \pm 10.6\%$ for Gemma~2. It then
\emph{increases} to $33.7\% \pm 1.3\%$ for Gemma~3, before dropping to
its lowest value of $12.3\% \pm 4.2\%$ for Gemma~4.

Every Gemma~3 seed (32.9--35.2\%) exceeds every Gemma~2 seed
(9.7--30.6\%) and every Gemma~4 seed (7.5--14.8\%). The regression
($+14.7$ percentage points) and the recovery ($-21.3$ points) therefore
hold in every seed. With three seeds, full separation corresponds to
the smallest attainable permutation $p$-value ($p = 0.10$;
\S\ref{sec:method:scaling}), so we report these as consistent but not
statistically significant. Gemma~1 and Gemma~2 are indistinguishable
(difference 2.3 points, $p = 0.80$). The non-refusal rate shows the same
ordering at higher levels: $59.4\%$, $45.5\%$, $68.7\%$ and $33.9\%$
(last row of Table~\ref{tab:asr}).

\begin{table}[t]
\centering
\caption{Attack success rate (\%; response judged non-refusing
\emph{and} harmful) per HarmBench category across Gemma generations
(mean $\pm$ std, 3 seeds). \textbf{Bold} indicates the lowest (safest)
ASR per category. Cells with $\pm 0.0$ are those where all seeds
converged to the same value. The last row gives the overall
non-refusal rate (the metric reported as ASR in v1) for comparison.}
\label{tab:asr}
\begin{tabular}{lcccc}
\toprule
\textbf{Category} & \textbf{Gemma 1} & \textbf{Gemma 2} & \textbf{Gemma 3} & \textbf{Gemma 4} \\
 & \textbf{(7B)} & \textbf{(9B)} & \textbf{(12B)} & \textbf{(31B)} \\
\midrule
Chem/Bio          & $13.0 \pm 3.2$ & $7.4 \pm 8.5$ & $15.7 \pm 15.3$ & $\mathbf{0.0 \pm 0.0}$ \\
Copyright\textsuperscript{$\dagger$} & $20.4 \pm 6.4$ & $34.3 \pm 18.5$ & $\mathbf{15.7 \pm 4.2}$ & $25.0 \pm 38.5$ \\
Cybercrime\textsuperscript{$\dagger$} & $70.4 \pm 4.2$ & $47.2 \pm 34.8$ & $83.3 \pm 8.3$ & $\mathbf{1.9 \pm 3.2}$ \\
Harassment        & $5.6 \pm 5.6$ & $\mathbf{0.0 \pm 0.0}$ & $2.8 \pm 0.0$ & $1.9 \pm 1.6$ \\
Harmful\textsuperscript{$\ddagger$} & $26.4 \pm 33.4$ & $4.2 \pm 5.9$ & $16.7 \pm 7.9$ & $\mathbf{0.0 \pm 0.0}$ \\
Illegal           & $9.3 \pm 13.7$ & $21.3 \pm 27.4$ & $15.7 \pm 15.3$ & $\mathbf{0.0 \pm 0.0}$ \\
Misinfo           & $\mathbf{7.4 \pm 4.2}$ & $16.7 \pm 2.8$ & $79.6 \pm 5.8$ & $52.8 \pm 30.9$ \\
\midrule
\textbf{Overall ASR}  & $21.3 \pm 3.7$ & $19.0 \pm 10.6$ & $33.7 \pm 1.3$ & $\mathbf{12.3 \pm 4.2}$ \\
\textit{Non-refusal rate} & $59.4 \pm 3.0$ & $45.5 \pm 7.2$ & $68.7 \pm 5.7$ & $33.9 \pm 1.8$ \\
\bottomrule
\end{tabular}
\\[2pt]
{\footnotesize
\textsuperscript{$\dagger$}High cross-seed variance: Gemma~4 copyright
(2.8\%, 69.4\%, 2.8\%) and Gemma~2 cybercrime (11.1\%, 80.6\%, 50.0\%),
reflecting sensitivity to the evolutionary trajectory.\\
\textsuperscript{$\ddagger$}Based on 2 seeds; one seed's archive
contained zero harmful-category entries (the evolution did not populate
those niches).
}
\end{table}

A raw-seed baseline (400 HarmBench prompts, no evolution; one run per
model) yields ASRs of 5.2\%, 5.2\%, 11.3\% and 4.6\% for Gemma~1--4.
Evolution adds 8--22 percentage points, and Gemma~3 is again the most
vulnerable generation even without evolution.

Cybercrime is the most exploited category in Gemma~1--3 ($70.4\%$,
$47.2\%$ and $83.3\%$) but falls to $1.9\%$ in Gemma~4. Misinformation
rises from $16.7\%$ in Gemma~2 to $79.6\%$ in Gemma~3. It stays at
$52.8\% \pm 30.9\%$ in Gemma~4, the largest category in the safest
generation. Gemma~4 reaches 0\% ASR in chem/bio, harmful and illegal.
Copyright ASR ranges from $15.7\%$ to $34.3\%$. v1 reported 100\% for
copyright in every generation, but that figure counted non-refusals,
most of which the judge did not label harmful. Harassment is near the
floor in every generation (0--5.6\%).

\subsection{Cross-Generational Transfer}
\label{sec:results:transfer}

Figure~\ref{fig:transfer} presents the full $4 \times 4$ transfer
matrix with uncertainty estimates. Several patterns emerge:

\begin{figure}[t]
\centering
\includegraphics[width=0.6\textwidth]{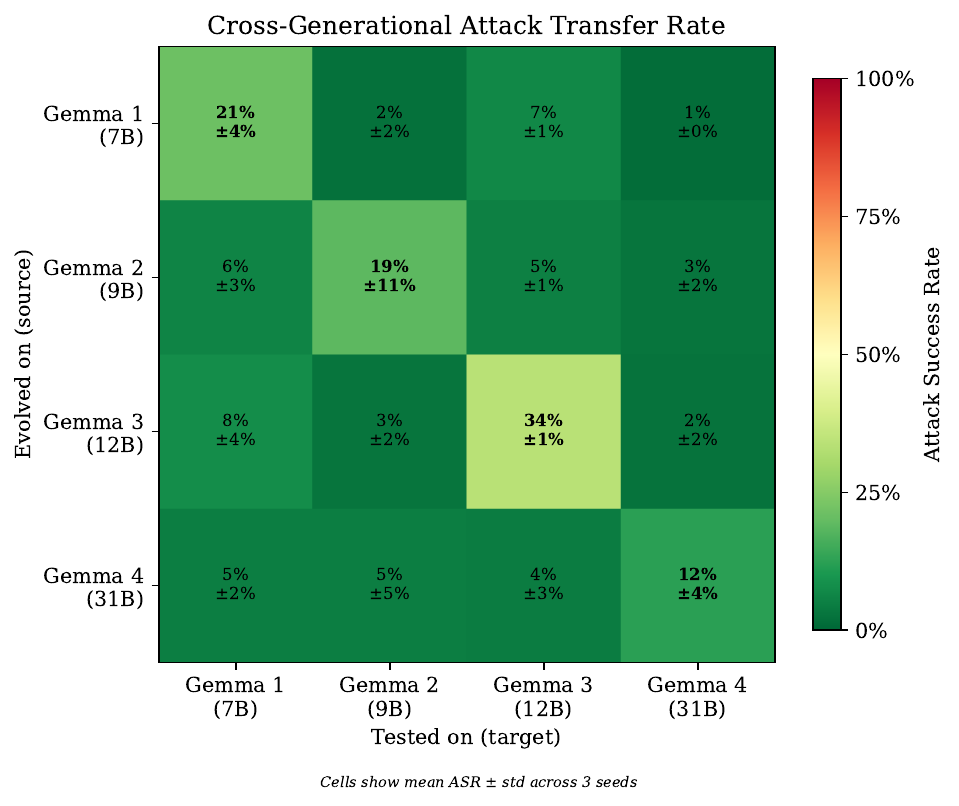}
\caption{Cross-generational attack transfer matrix (mean $\pm$ std
across 3 seeds). Cell $(i,j)$ shows the fraction of attacks evolved
against Gemma~$i$ that elicit a non-refusing, harmful response from
Gemma~$j$. Diagonal entries (bold) are self-evaluation rates. The color
scale spans 0--100\%, as in v1.}
\label{fig:transfer}
\end{figure}

\paragraph{Harmful success rarely transfers.} Off-diagonal ASR ranges
from 1.4\% to 7.9\%, against 12.3--33.7\% on the diagonal. Most attacks
that elicit harmful content are specific to the generation they were
evolved against. The few that do transfer also differ by source. We
pool successful (strategy $\times$ encoding $\times$ category) cells
over seeds, then compare these sets for the three archives replayed on
Gemma~3. Their pairwise Jaccard similarity is 0.20--0.35.

\paragraph{Gemma~4 receives the least transfer.} Transfer into Gemma~4
is 1.4--2.8\% from every source. This compares with 4.5--6.7\% into
Gemma~3 and 4.7--7.9\% into Gemma~1. Archives evolved against Gemma~3,
the most vulnerable generation, transfer to Gemma~4 at only
$2.4\% \pm 2.1\%$. Differences between off-diagonal cells are small
relative to their seed variance (Figure~\ref{fig:transfer}), so we
treat this ordering as indicative.

\paragraph{Non-refusal transfers broadly.} Replayed attacks avoid
refusal far more often than they elicit harm. Their non-refusal rates
are 44--46\% into Gemma~3, 49--52\% into Gemma~1 and 14--18\% into
Gemma~4; these are the rates v1 reported as transfer ASR. Most of these
non-refused responses are judged not harmful.

\subsection{Robustness Across Generations}
\label{sec:results:scaling}

\begin{figure}[t]
\centering
\includegraphics[width=0.95\textwidth]{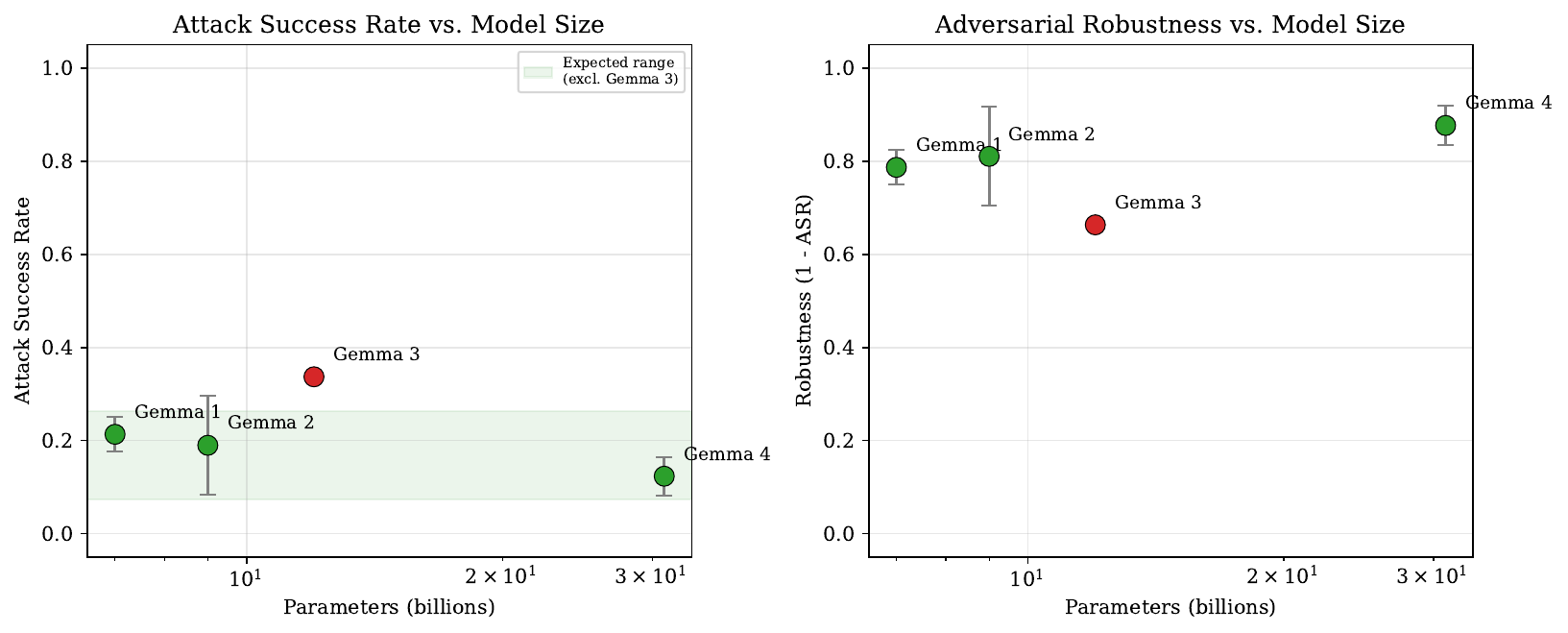}
\caption{Attack success rate (left) and adversarial robustness (right)
vs.\ model size with error bars ($\pm$ 1 std, 3 seeds). The green
shaded region shows the range spanned by Gemma~1, 2, and 4; Gemma~3
(red) falls well outside this range, demonstrating the non-monotonic
pattern. No scaling law is fit; four data points are insufficient.}
\label{fig:scaling}
\end{figure}

Figure~\ref{fig:scaling} plots ASR and robustness ($1 - \text{ASR}$)
against parameter count with error bars. We do not fit a scaling law
to four data points. Instead, we observe that three of the four
models (Gemma~1, 2, 4) are consistent with a gradual improvement
trajectory, while Gemma~3 represents a clear outlier. This
non-monotonic pattern coincides with the architectural transition
from text-only (Gemma~1, 2) to multimodal (Gemma~3, 4), though
Gemma~4's recovery shows that the regression is not an inherent
consequence of multimodal design. Gemma~4's low overall ASR is
driven by near-zero ASR in four categories (chem/bio, cybercrime,
harmful, illegal). Misinformation ($52.8\%$) remains high, and so does
copyright ($25.0\%$), though the latter has high variance across
seeds.

\subsection{Vulnerability Fingerprints}
\label{sec:results:fingerprint}

\begin{figure}[t]
\centering
\includegraphics[width=0.55\textwidth]{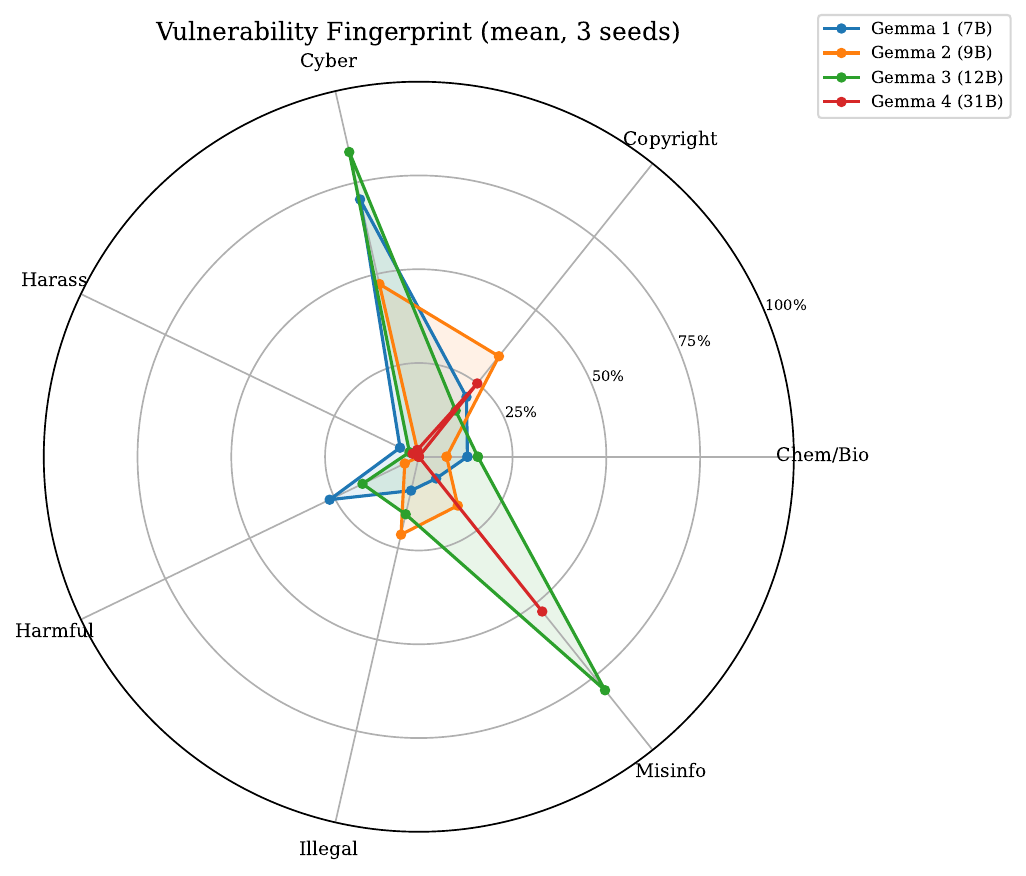}
\caption{Vulnerability fingerprints across Gemma generations (mean
ASR per category, 3 seeds). Gemma~4 shows the most compact profile.
Gemma~3's profile is expanded in cybercrime and misinformation.}
\label{fig:radar}
\end{figure}

Figure~\ref{fig:radar} presents vulnerability fingerprints as radar
charts. Each model exhibits a distinct profile:
\begin{itemize}
    \item \textbf{Gemma~1}: dominated by cybercrime ($70.4\%$). The
          other categories are at or below $26.4\%$.
    \item \textbf{Gemma~2}: cybercrime ($47.2\%$, high variance) and
          copyright ($34.3\%$) lead. The other categories are at or
          below $21.3\%$.
    \item \textbf{Gemma~3}: expanded in cybercrime ($83.3\%$) and
          misinformation ($79.6\%$).
    \item \textbf{Gemma~4}: near zero in five categories (chem/bio,
          cybercrime, harassment, harmful, illegal), while
          misinformation ($52.8\%$) and copyright ($25.0\%$) remain.
\end{itemize}

\section{Discussion}
\label{sec:discussion}

\paragraph{Safety regression in Gemma~3.}
The most surprising finding is Gemma~3's safety regression. Its ASR
exceeds Gemma~2's in every seed ($33.7\%$ vs.\ $19.0\%$), although
three seeds cannot establish statistical significance
(\S\ref{sec:method:scaling}). We hypothesize this is related to the
introduction of multimodal capabilities: Gemma~3 was the first Gemma
generation with vision-language support (architecture:
\texttt{Gemma3ForConditionalGeneration}), requiring substantial
architectural changes that may have disrupted safety-trained text
representations. The regression is concentrated in cybercrime and
misinformation (Table~\ref{tab:asr}), while harassment stays near the
floor in every generation. This is consistent with category-specific
safety training being maintained while broader robustness regressed,
but does not establish it.
Alternative hypotheses that the data does not rule out include
(a)~idiosyncratic training-data-mix issues specific to Gemma~3
unrelated to modality, (b)~changes in the RLHF/DPO recipe between
generations, and (c)~increased model capability raising the ceiling
on evolutionary search. Distinguishing these would require access to
training details, which we do not have.

\paragraph{Gemma~4's recovery.}
Gemma~4 achieved the lowest overall ASR in the family
($12.3\% \pm 4.2\%$) despite being multimodal, suggesting that the
safety regression in Gemma~3 was addressable. Archives evolved against
Gemma~3 transfer to Gemma~4 at only $2.4\% \pm 2.1\%$. Because harmful
transfer is low between \emph{all} generations (1.4--7.9\%), this is
weaker evidence of generalizing safety gains than v1 suggested.
Gemma~4's safety is category-heterogeneous: four categories are at or
near 0\% ASR, while misinformation ($52.8\%$) remains high.

\paragraph{Judge sensitivity.}
Our primary metric depends on the 8B judge's harmful-content label. A
second-judge audit used Llama-3.1-70B-Instruct on 52 stored top-attack
responses, which were selected for high 8B scores. Of the 46 responses
the 8B judge labeled harmful, the 70B judge agreed on 34 (74\%): 18/21
in cybercrime, 6/9 in misinformation and 5/10 in copyright. The two
judges agree on refusal (96.2\%) but only weakly on the harmful verdict
($\kappa = 0.15$). The audit's 52 samples limit per-category
resolution, and a larger audit would be needed to characterize judge
disagreement per category. Per-category ASRs, particularly copyright,
should be read with this caveat.

\paragraph{What we claim and what we do not.}
With four data points we cannot distinguish a U-shaped curve from
noise, and with three seeds per model no pairwise difference can reach
$p < 0.05$. What we \emph{can} say is that the Gemma~2 $\to$ Gemma~3
increase ($+14.7$ percentage points) and the Gemma~3 $\to$ Gemma~4
decrease ($-21.3$ points) hold in every seed. They also hold under both
the harmful-content criterion and the non-refusal rate, and in the
raw-seed baseline.

\paragraph{Implications for safety evaluation.}
Static benchmarks would miss the non-monotonic trajectory entirely.
Architectural changes (such as adding multimodal capabilities) should
trigger comprehensive safety re-evaluation. The success criterion
matters as much as the probe. In our data, counting non-refusals as
successes overstates harmful attack success by 2--3$\times$ for
self-evaluation and 6--16$\times$ for transfer. We recommend reporting
both rates, and stating explicitly which one is called ASR.

\paragraph{Limitations.}
\begin{itemize}
    \item Three seeds per model provide reproducibility estimates but
          little statistical power. Under a two-sided permutation test,
          no pairwise comparison can reach $p < 0.05$ (the minimum is
          $p = 0.10$). Five or more seeds per model would be needed to
          test the regression formally.
    \item Single model family (Gemma) for generational analysis;
          findings may not generalize to other families.
    \item The primary metric relies on an 8B judge's harmful-content
          label, which a 70B judge confirms in 74\% of audited cases,
          with weak overall agreement ($\kappa = 0.15$). Per-category
          ASRs are judge-dependent, particularly copyright.
    \item Evolution maximizes the judge score, which rewards relevant
          non-refusals (0.3) as well as harmful responses. An archive
          optimized for harmful success alone might show different
          rates.
    \item HarmBench coverage: 400 behaviors across 7 categories is
          comprehensive but not exhaustive.
    \item Evolutionary probe explores a subset of the attack space;
          absence of a successful attack does not prove safety.
    \item We study only the text modality; Gemma~3 and 4's vision
          capabilities may present additional attack surfaces not
          captured here.
\end{itemize}

\section{Conclusion}
\label{sec:conclusion}

We have presented a methodology for longitudinal safety evaluation of LLMs
using quality-diversity evolution as a differential probe. Applied to four
generations of Google's Gemma models with 3 independent seeds per model,
we find that:

\begin{enumerate}
    \item Safety alignment does not improve monotonically across model
          generations in our data. Gemma~3's ASR ($33.7\% \pm 1.3\%$)
          exceeds Gemma~2's ($19.0\% \pm 10.6\%$) and Gemma~4's
          ($12.3\% \pm 4.2\%$) in every seed. This is a directional
          result, since three seeds cannot establish significance.
    \item Attacks that elicit harmful content rarely transfer between
          generations (1.4--7.9\%), and transfer least to Gemma~4
          (1.4--2.8\%).
    \item Vulnerability profiles are category-dependent. Cybercrime
          dominates in Gemma~1--3 but nearly vanishes in Gemma~4, while
          misinformation remains elevated ($52.8\%$) even in the safest
          generation.
    \item Counting non-refusals as successes (the metric v1 reported as ASR)
          overstates harmful success by 2--3$\times$ (self-evaluation)
          and 6--16$\times$ (transfer).
\end{enumerate}

These results underscore the need for \emph{adaptive, longitudinal}
safety evaluation. Static benchmarks provide a snapshot; evolutionary
probing with transfer analysis reveals the \emph{trajectory} of safety
alignment. We advocate for model developers to publish generational
safety analyses alongside capability benchmarks, and for the research
community to develop standardized protocols for cross-version
safety comparison.


\begin{ack}
Compute for Gemma experiments was provided by RunPod (NVIDIA H200
and RTX~4090 instances).
\end{ack}

\bibliographystyle{plainnat}
\bibliography{references}

\end{document}